\documentclass[12pt]{iopart}

\newcommand{\bra}[1]{\mbox{$\left\langle#1\right|$}}
\newcommand{\ket}[1]{\mbox{$\left|#1\right\rangle$}}
\newcommand{\inpro}[2]{\mbox{$\left\langle#1|#2\right\rangle$}}

\newcommand{\proj}[1]{\mbox{$\ket{#1}\!\bra{#1}$}}

\usepackage{iopams} 
\usepackage{bm}
\usepackage{latexsym}
\usepackage{amssymb}
\usepackage{theorem}
\usepackage[dvipdfmx]{graphicx}
\usepackage{color}
\usepackage{ulem}

{\theorembodyfont{\normalfont}
\theoremheaderfont{\normalfont\bfseries}
\newtheorem{thm}{Theorem}[section]
\newtheorem{dfn}[thm]{Definition}
\newtheorem{lmm}[thm]{Lemma}

\newtheorem{prp}[thm]{Property}
}

{\theorembodyfont{\normalfont}
\theoremheaderfont{\normalfont\bfseries}
\newtheorem{prf}{Proof.}
\newtheorem{lmmA1}[prf]{Lemma A.1}
\newtheorem{lmmA2}[prf]{Lemma A.2}
\newtheorem{thmA3}[prf]{Theorem A.3}
\newtheorem{dfnB1}[prf]{Definition B.1}
\newtheorem{dfnB2}[prf]{Definition B.2}
}

\begin{document}

\title[The chain rule implies Tsirelson's bound]{The chain rule implies Tsirelson's bound: an approach from generalized mutual information}

\author{Eyuri Wakakuwa$^{1}$ and Mio Murao$^{1,2}$}

\address{$^1$Department of Physics, Graduate School of Science, The University of Tokyo, Tokyo 113-0033, Japan}
\address{$^2$Institute for Nano Quantum Information Electronics, The University of Tokyo, Tokyo 113-0033, Japan}
\ead{wakakuwa@eve.phys.s.u-tokyo.ac.jp}

\begin{abstract}
{In order t}o analyze {an}  information theoretical derivation of Tsirelson's bound based on information causality, we introduce a generalized mutual information (GMI){, defined} as the optimal coding rate of a channel with classical inputs and general probabilistic outputs. In {the case where the outputs are quantum}, the GMI coincides with the quantum mutual information. In general, the GMI does not necessarily satisfy the chain rule. We prove that Tsirelson's bound can be derived {by imposing} the chain rule {on} the GMI. {W}e formulate a principle, which we call the {{\it no-supersignalling condition}}, {which states} that the assistance of nonlocal correlations does not increase the capability of classical communication. We prove that this condition is equivalent to the no-signalling condition. As a result, we show that {Tsirelson's bound is implied by the nonpositivity of the quantitative difference between information causality and no-supersignalling}.
\end{abstract}


\maketitle

\section{Introduction}

One of the most counterintuitive phenomena that quantum mechanics predicts is nonlocality. {The statistics of the outcomes of {m}easurements performed {on an entangled state} at two space-like {separated} points can exhibit strong correlations that cannot be described within the framework of local realism.  This {can be formulated in terms of the violation of Bell inequalities} \cite{bell}.   {On the other hand, it is also known that} quantum correlations still {satisfy} the no-signalling condition, i.e., they cannot be used for superluminal communication, which is prohibited by {s}pecial relativity. {The amount that quantum mechanics can violate} the Clauser-Horne-Shimony-Holt (CHSH) inequality \cite{chsh} is limited by Tsirelson's bound \cite{tsirelson}. In a seminal paper \cite{prbox}, Popescu and Rohrlich showed that Tsirelson's bound is strictly lower than the limit imposed by the no-signalling condition alone. This result raise{s} the question {of} why the strength of nonlocality is limited to Tsirelson's bound in the quantum world. If we could find an operational principle rather than mathematical one to answer this question, it would help us better understand why quantum mechanics is the way it is \cite{cc1,cc2,cc3}.

From an information theoretical point of view, it is natural to ask if {superstrong?} nonlocality, i.e., nonlocal correlations exceeding Tsirelson's bound, can be used to increase the capability of {classical communication} \cite{ic1}. Suppose that Alice is trying to send classical information to distant Bob {with} the assistance of nonlocal correlations shared in advance. The no-signalling condition implies that, if no classical communication from Alice to Bob is performed, Bob's information gain is zero bits. In other words, zero bits of classical communication can produce no more than zero bits of classical information gain {for} the receiver. On the other hand, the no-signalling condition does not eliminate the {possibility that $m>0$} bits of classical communication produces more than $m$ bits of classical information gain {for} the receiver. Whether such an implausible situation can occur would be depending on the strength of nonlocal correlations. {In particular}, one {might} expect that Tsirelson's bound {could be} derived from the impossibility {o}f such a situation.

Motivated by the foregoing consideration{s}, {\it information causality} {has been} proposed as an answer to the question \cite{ic1}. Information causality is the condition that {\it {i}n bipartite nonlocality-assisted random access coding protocols, the receiver's {total} information gain cannot be greater than the amount of classical communication allowed in the protoco{l}}. This condition is never violated in classical {or} quantum theory, whereas it is violated in all ``supernonlocal''  theories, i.e., theories that predict supernonlocal correlations \cite{ic1}. It implies that Tsirelson's bound is derived from this purely information theoretical principle. Thus information causality is regarded as one of the basic informational principles at the foundations of quantum mechanics.

In \cite{ic1}, it is proved that information causality is never violated in any no-signalling theory {in which} we can define {a {\it mutual information}} satisfying {five particular} propertie{s. This} implies that in supernonlocal theories, we cannot define a function {like} the mutual information that satisfies all five{.} On the other hand, {both the classical and quantum mutual information} satisfy all of the five properties. {It is therefore} natural to ask another question: which of the five properties is lost in supernonlocal theories? We address this question {to better understand} the informational features of supernonlocal theories in comparison with quantum theory.

In order to answer this question, we need to define {a generalization of the quantum mutual information} that is applicable to general probabilistic theories. Several investigations have been made along this line. In \cite{barnum,short}, a generalized entropy $H$ is define{d}, and then a mutual information is defined in terms of th{is} by $I(A:B):=H(A)+H(B)-H(A,B)$. {U}sing this mutual information, it is proved that the data processing inequality is not satisfied in supernonlocal theories. Similar results are obtained in \cite{oscar,alsafi}. However, the definitions of the entropies in their approaches are mathematical, and do not have clear operational meanings. Note that in classical and quantum information theory, the operational meaning of {entropy and mutual} information is given by the source coding and channel coding theorems. In \cite{short}, a coding theorem analogous to Schumacher's quantum coding theorem \cite{schu0} is investigated using generalized entropy. However, their consideration is only applicable under several restrictions. As discussed in \cite{barnum}, we need to seek {generalization}s based on the analysis of data compression or channel capacit{y}. Such an approach is also {studied} in \cite{gray}.

{Motivated by these discussions}, we introduce an operational definition of {g}eneralized mutual information (GMI) that is applicable to any general probabilistic theor{y}. This is a generalization of the quantum mutual information between a classical system and a quantum system. Unlike the previous entropic approaches, we directly address the mutual information. The generalization is based on the channel coding theorem.  Thus the GMI inherently has an operational meaning as {a} transmission rate of classical information. Our definition does not require {mathematical notions} such as {state space or} fine-grained measurement. The GMI is defined between a classical system and a general probabilistic system {-- it} is not applicable to two general probabilistic systems, but it is sufficient for analyzing the situation describing information causality. The GMI satisfies four of the five properties of the mutual information{, the exception being} the chain rule. {We will show that} violation of Tsirelson's bound {implies v}iolation of the chain rule of the GMI. 

{U}sing the GMI, we further investigate the derivation of Tsirelson's bound in terms of information causality. We formulate a principle, which we call the {\it no-supersignalling condition}, {stating} that the assistance of nonlocal correlations does not increase the capability of classical communication. We prove that this condition is equivalent to the no-signalling condition, and thus it is different from information causality. This result is similar to the result obtained in \cite{short}, but now become{s} operationally supported. It implies that Tsirelson's bound is not derived from the condition that ``$m$ bits of classical communication cannot produce more than $m$ bits of information gain''. We show that Tsirelson's bound is derived from the nonpositivity of the quantitative difference between information causality and no-supersignalling. Our results indicate that the chain rule of the GMI imposes a strong restriction on the underlying physical theory. As an example of this fact, we show that we can derive a bound on the state space of {\it one} gbit from the chain rule. 

This paper is organized as follows.  In Section \ref{sec:gpt}, we introduce a minimal framework for general probabilistic theories. In Section \ref{sec:ic}, we give a brief review of information causality. In Section \ref{sec:mi}, we define the generalized mutual information, and show that Tsirelson's bound is derived from the chain rule. In Section \ref{sec:quantum}, we prove that the GMI is a generalization of the quantum mutual information. In Section \ref{sec:cr}, we formulate the no-supersignalling condition, and prove that the condition is equivalent to the no-signalling condition. In Section \ref{sec:nec}, we clarify the relation among no-supersignalling, information causality and Tsirelson's bound. In Section \ref{sec:ex}, we show that we can limit the state space of one gbit by assuming the chain rule. We conclude with a summary and discussio{n} in Section \ref{sec:conclusion}.

\section{General probabilistic theories}
\label{sec:gpt}
In this section we introduce a minimal framework for general probabilistic theories based on \cite{short,nobroad}.

We associate a set of allowed {\it states} ${\mathcal S}_S$ with each physical system $S$. We assume that any probabilistic mixture of states is also a state, i.e., if $\phi_1\in{\mathcal S}_S$ and $\phi_2\in{\mathcal S}_S$ then $\phi_{\rm mix}=p\phi_1+(1-p)\phi_2\in{\mathcal S}_S$, where $p\phi_1+(1-p)\phi_2$ denotes the state that is a mixture of $\phi_1$ with probability $p$ and $\phi_2$ with probability $1-p$.

We also associate a set of allowed {\it measurements} ${\mathcal M}_S$ with each system $S$. A set of outcomes ${\mathcal R}_e$ is associated with each measurement $e\in{\mathcal M}_S$. The state determines the probability of obtaining an outcome $r\in{\mathcal R}_e$ when a measurement $e\in{\mathcal M}_S$ is performed on the system $S$. Thus we associate each outcome $r\in{\mathcal R}_e$ with a functional $e_r:{\mathcal S}\rightarrow[0, 1]$, such that $e_r(\phi)$ is the probability of obtaining outcome $r$ when a measurement $e$ is performed on a system in the state $\phi$. Such a functional is called an {\it effect}. In order that the statistics of measurements on mixed states fits into our intuition, we require the linearity of each effect, i.e., $e_r(\phi_{\rm mix})=pe_r(\phi_1)+(1-p)e_r(\phi_2)$.

It may be possible to perform {\it transformations} on a system. A transformation on the system $S$ is described by a map ${\mathcal E}:{\mathcal S}_S\rightarrow{\mathcal S}_{S'}$, where $S'$ {denotes} the output system. We assume the linearity of transformations, i.e., ${\mathcal E}(\phi_{\rm mix})=p{\mathcal E}(\phi_1)+(1-p){\mathcal E}(\phi_2)$. A measurement $e\in{\mathcal M}_S$ is represented by a transformation ${\mathcal E}_{\rm M}:{\mathcal S}_S\rightarrow{\mathcal S}_{T_S}$, where $T_S$ represents a classical system corresponding to the register of the measurement outcome. We assume that the composition of two allowed transformations is also an allowed transformation, and that any allowed transformation followed by an allowed measurement is an allowed measurement.

We assume that a composition of two systems is also a system. If we have two systems $A$ and $B$, we can consider a composite system $AB$ which has its own set of allowed states ${\mathcal S}_{AB}$ and that of allowed measurements ${\mathcal M}_{AB}$. Suppose that measurements $e_A\in{\mathcal M}_{A}$ and $e_B\in{\mathcal M}_{B}$ are performed on the system $A$ and $B$, respectively. Such a measurement is called a {\it product measurement} and is included in ${\mathcal M}_{AB}$. We assume that a global state $\psi\in{\mathcal S}_{AB}$ determines a joint probability for each pair of effects $(e_{A,r},e_{B,r'})$. We may also assume that the global state is  uniquely specified if the joint probabilities for all pairs of effects $(e_{A,r},e_{B,r'})$ are specified. Such an assumption is called the {\it global state assumption}. However, it is known that there exists general probabilistic theories which do not fit into this assumption, such as quantum theory in a real Hilbert space. The arguments {presented in the following sections} of this paper are developed under the global state assumption, although the main results are valid without this assumption. The generalization for theories without this assumption is given in \ref{sec:gsa}.

\section{Review of information causality}
\label{sec:ic}
Information causality{, introduced in \cite{ic1},} is the principle that {\it the total amount of classical information gain that the receiver can obtain in a bipartite nonlocality-assisted random access coding protocol cannot be greater than the amount of classical communication that is allowed in the protocol}. Suppose that a string of $n$ random and independent bits $\vec{X}=X_1,\cdots,X_n$ is given to Alice, and a random number $k\in\{1,\cdots,n\}$ is given to distant Bob. The task is for Bob to correctly guess $X_k$ under the condition that they can use a resource of shared correlations and a $m$ bit one way classical communication from Alice to Bob (see Figure \ref{icsit}). To accomplish this task, Alice first performs a measurement on her part of the resource (denoted by $A$ in the figure), depending on $\vec{X}$. She then constructs a $m$ bit message $\vec{M}$ from $\vec{X}$ and the measurement outcome, and sends it to Bob. Bob, after receiving $\vec{M}$, performs a measurement on his part of the resource (denoted by $B$ in the figure), depending on $\vec{M}$ and $k$. From the outcome of the measurement he computes his guess $G_k$ for $X_k$. The efficiency of the protocol {is} quantified by
\begin{eqnarray}
J:=\sum_{k=1}^{n}I_C(X_k:G_k)\;,
\label{eq:jdef}
\end{eqnarray}
where $I_C(X_k:G_k)$ is the classical (Shannon) mutual information between $X_k$ and $G_k$. Information causality is the condition that, whatever strategy they take and whatever resource of shared correlation allowed in the theory they use,
\begin{eqnarray}
J\leq m\;
\label{eq:icdef}
\end{eqnarray}
must hold for all $m\geq0$. The derivation of Tsirelson's bound in terms of information causality consists of the following two theorems that are proved in \cite{ic1}. 

\begin{thm}
If we can define a function $I(A:B)$ satisfying the following five properties in the general probabilistic theory, $J\leq m$ holds for all $m\geq0$. The properties are
\begin{itemize}
\item {\it Symmetry} : $I(A:B)=I(B:A)$ for any systems $A$ and $B$.
\item {\it Nonnegativity} : $I(A:B)\geq0$ for any systems $A$ and $B$.
\item {\it Consistency} : If both systems $A$ and $B$ are in classical states, $I(A:B)$ coincides with the classical mutual information.
\item {\it Data Processing Inequality} : Under any local transformation that maps states of system $B$ into states of another system $B'$ without post-selection, $I(A:B)\geq I(A:B')$.
\item {\it Chain Rule} : For any systems $A$, $B$ and $C$, the {\it conditional mutual information} defined by $I(A:B|C):=I(A:B,C)-I(A:C)$ is symmetric in $A$ and $B$.
\end{itemize}
\label{thm:icsat}
\end{thm}
\begin{thm}
If there exists a nonlocal correlation exceeding Tsirelson's bound, we can construct a nonlocality-assisted communication protocol by which $J>m$ is achieved.
\label{thm:icvio}
\end{thm}

Theorem \ref{thm:icsat} guarantees that both classical and quantum theory satisfy information causality. Theorem \ref{thm:icvio} implies that information causality is violated in all supernonlocal theories. These two theorems imply that, in any supernonlocal theory, we cannot define a function of the mutual information that satisfies all five properties.

\begin{figure}[t]
\centerline{\includegraphics[scale=0.55]{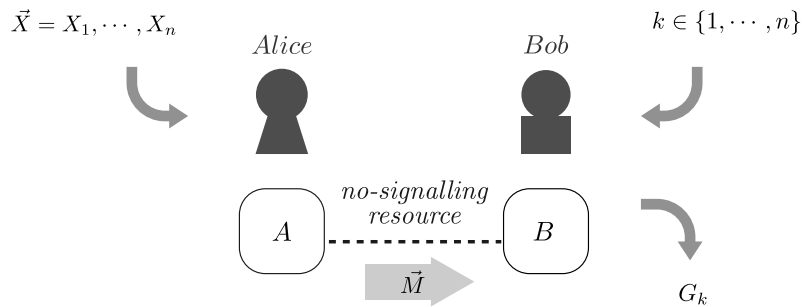}} 
\caption{Nonlocality-assisted random access coding. The task is for Bob to correctly guess $X_k$, where $k$ is a random number unknown to Alice.}
\label{icsit}
\end{figure}

\section{Generalized mutual information}
\label{sec:mi}
Suppose that there are a classical system $X$ and a system $S$ that is described by a general probabilistic theory. The states of $X$ are labeled by a finite alphabet ${\mathcal X}$. For each state $x$ of $X$, the corresponding state of $S$ denoted by $\phi_x$ is determined. The state of the composite system $XS$ is determined by a probability distribution $p(x)={\rm Pr}(X=x)$, which represents the probability that the system $X$ is in the state $x$, and the corresponding state $\phi_x$ of $S$. Thus the state of the composite system $XS$ is identified with an ensemble $\{p(x),\phi_x\}_{x\in{\mathcal X}}$. To define a generalized mutual information $I_G(X:S)$ between the system $X$ and the system $S$ in the state $\{p(x),\phi_x\}_{x\in{\mathcal X}}$, we analyze the classical information capacity of a channel that outputs the system $S$ in the state $\phi_x$ according to the input $X=x$ (Figure \ref{fig:channel1}). As usually considered in information theory, the sender Alice, who has access to $X$, tries to send classical information to the receiver Bob, who has access to $S$, by using the channel many times. Suppose that they use $l$ identical and independent copies of this channel. Let $X_1,\cdots,X_l$ be the inputs of the $l$ channels and $S_1,\cdots,S_l$ be the corresponding output systems.

\begin{figure}[t]
\centerline{\includegraphics[scale=0.44]{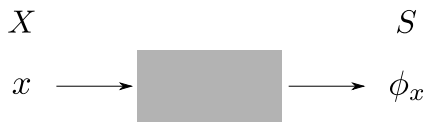}} 
\caption{The channel {defining} the mutual information between the system $X$ and the system $S$. It has a classical system as the input system and a general probabilistic system as the output system.}
\label{fig:channel1}
\end{figure}

Alice's encoding scheme is determined by a codebook. Let $w\in\{1,\cdots,N\}$ be a message that Alice tries to communicate, and the codeword $x^l(w)=x_1(w)\cdots x_l(w)$ be the corresponding input sequence to the channels. The codebook $\mathcal C$ is defined as the list of the codewords for all messages by
\begin{eqnarray}
{\mathcal C}:=\left[
\begin{array}{ccc}
  x_1(1) & \cdots & x_l(1) \\
  \vdots & \ddots & \vdots \\
  x_1(N) & \cdots & x_l(N) 
\end{array}
\right]\;.
\end{eqnarray}
The letter frequency $f(x)$ for the codebook is defined by
\begin{eqnarray}
f(x):=\frac{|\{(k,w)|x_k(w)=x,1\leq k\leq l, 1\leq w\leq N\}|}{lN}\qquad(x\in{\mathcal X})\;.
\end{eqnarray}
For a given probability distribution $\{p(x)\}_{x\in\mathcal X}$, the tolerance $\tau$ of the code is defined by 
\begin{eqnarray}
\tau:=\max_{x\in\mathcal X}|p(x)-f(x)|\;.
\end{eqnarray}

By performing a decoding measurement on the output systems $S_1,\cdots,S_l$, Bob tries to guess what the original message $w$ is. Let ${\mathcal D}$ denote the decoding measurement. Note that, in general, the decoding measurement is not one in which Bob performs a measurement on each of $S_1,\cdots,S_l$ individually, but one in which the whole of the composite system $S_1\cdots S_l$ is subjected to a measurement. Let $W$, $\hat W$ be Alice's original message and Bob's decoding outcome, respectively. The average error probability $P_e$ is defined by
\begin{eqnarray}
P_e:=\frac{1}{N}\sum_{u=1}^{N}{\rm Pr}(\hat W\neq u|W=u)\;.
\end{eqnarray} 
The pair of the codebook $\mathcal C$ and the decoding measurement ${\mathcal D}$ is called an $(N,l)$ code. The ratio $\log{N}/l$ is called the rate of the code, and represents how many bits of classical information is transmitted per use of the channel.

\begin{dfn}
A rate $R$ is said to be achievable with $p(x)$ if there exists a sequence of $(2^{lR},l)$ codes $({\mathcal C}^{(l)},{\mathcal D}^{(l)})$ such that 
\begin{enumerate}
\item $P_e^{(l)}\rightarrow0$ when $l\rightarrow\infty$,
\item $\tau^{(l)}\rightarrow 0$ when $l\rightarrow\infty$.
\end{enumerate}
\end{dfn}

\begin{dfn}
The mutual information between a classical system $X$ and a general probabilistic system $S$, denoted by $I_G(X:S)$, is the function which satisfies the condition that
\begin{enumerate}
\item A rate $R$ is achievable with $p(x)$ if $R<I_G(X:S)$,
\item A rate $R$ is achievable with $p(x)$ only if $R\leq I_G(X:S)$.
\end{enumerate}
We also define $I_G(S:X)$ by $I_G(S:X):=I_G(X:S)$.
\label{def:igdef}
\end{dfn}

\begin{thm}
$I_G(X:S)$ exists and satisfies $I_G(X:S)\leq H(X)$. Here, $H(X)$ is the Shannon entropy of the system $X$ defined by $H(X):=-\sum_{x\in\mathcal X}p(x)\log{p(x)}$.
\label{thm:existence}
\end{thm}

\begin{prf}
First we prove the existence of $R^*:=\sup{\{R|R\textrm{ is achievable with }p(x)\}}$. Consider a $(2^{lR},l)$ code and suppose that Alice's message $W=1,\cdots,2^{lR}$ is uniformly distributed. Let $I'$, $H'$ be the mutual information and the entropy when the input sequence is the codeword corresponding to the uniformly distributed message $W$. By Fano's inequality, we have
\begin{eqnarray}
H'(W|\hat{W})\leq P_e^{(l)}lR+1
\end{eqnarray}
where $P_e^{(l)}=P(W\neq \hat{W})$. Thus
\begin{eqnarray}
lR=H'(W)&=I'(W:\hat{W})+H'(W|\hat{W})\nonumber\\
		&\leq I'(X^l:{\hat W})+P_e^{(l)}lR+1\nonumber\\
		&\leq H'(X^l)+P_e^{(l)}lR+1\;.
\label{eq:iexi1}
\end{eqnarray}
Here, we use the data processing inequality in the first inequality. By introducing a classical variable $K$ that indicates $k$ with the probability distribution $P(K=k)=1/l$, we also have
\begin{eqnarray}
		H'(X^l)\leq\sum_{k=1}^{l}H'(X_k)=lH'(X|K)\leq lH'(X)\;,
\label{eq:iexi2}
\end{eqnarray}
where $X$ is a random variable defined by ${\rm Pr}(X=x_k(w))=2^{-lR}/l$. From (\ref{eq:iexi1}) and (\ref{eq:iexi2}), we obtain
\begin{eqnarray}
P_e^{(l)}\geq1-\frac{H'(X)}{R}-\frac{1}{lR}\;.
\end{eqnarray}
If $R$ is achievable with $p(x)$, there exists a sequence of $(2^{lR},l)$ codes satisfying $P_e^{(l)}\rightarrow0$ and $H'(X)\rightarrow H(X)$ when $l\rightarrow \infty$. Thus $R\leq H(X)$. Hence $R^*$ exists and satisfies $R^*\leq H(X)$.

Next we prove that any rate $R<R^*$ is also achievable with $p(x)$. Let  $\{({\mathcal C}^{*(l)},{\mathcal D}^{*(l)})\}_l$ be a sequence of $(2^{lR^*},l)$ codes that satisfies $P_e^{*(l)}\rightarrow0$ and $\tau^{*(l)}\rightarrow 0$. For arbitrary $0\leq\lambda<1$, define another codebook ${\mathcal C}^{(l)}$ by using ${\mathcal C}^{*(\lambda l)}$ for the first $\lambda l$ codeletters and by choosing the last $(1-\lambda)l$ codeletters arbitrarily so that the total tolerance is sufficiently small. Also define the corresponding decoding measurement ${\mathcal D}^{(l)}$ as the measurement in which the output system $S_1\cdots S_{\lambda l}$ is subjected to the decoding measurement ${\mathcal D}^{*(l)}$ and the output systems $S_{\lambda l+1},\cdots,S_{l}$ are ignored. The code sequence $\{({\mathcal C}^{(l)},{\mathcal D}^{(l)})\}_l$ constructed in this way is a sequence of $(2^{l\lambda R^*},l)$ codes that satisfies $P_e^{(l)}\rightarrow0$ and $\tau^{(l)}\rightarrow 0$. Thus $R=\lambda R^*$ is achievable with $p(x)$. Hence we obtain $R^*=I_G(X:S)$.
 \hfill{$\square$}
\end{prf}

Note that $I_G(X:S)$ is a function of the state $\Gamma:=\{p(x),\phi_x\}_{x\in{\mathcal X}}$ of the composite system $XS$. To emphasize this, we sometimes use the notation $I_G(X:S)_{\Gamma}$. Since $R=0$ is always achievable, $I_G(X:S)$ is nonnegative. Shannon's noisy channel coding theorem guarantees that $I_G(X:S)$ coincides with the classical mutual information $I_C(X:S)$ if $S$ is a classical system \cite{thomas}. The generalized mutual information satisfies the data processing inequality as follows.

\begin{prp}
Let ${\mathcal E}_{S\rightarrow S'}$ be any local transformation that maps states of a general probabilistic system $S$ into states of another general probabilistic system $S'$. If ${\mathcal E}_{S\rightarrow S'}$ contains no post-selection, the generalized mutual information does not increase under this transformation, i.e., $I_G(X:S)\geq I_G(X:S')$. Similarly, $I_G(X:S)\geq I_G(X':S)$ under any local transformation ${\mathcal E}_{X\rightarrow X'}$ that maps states of a classical system $X$ into states of another classical system $X'$ without post-selection. 
\label{prp:dpi}
\end{prp}
\begin{prf}
Here we only prove the former part. For the latter part, see Appendix A. Consider two channels, {channel I and} channel II (see Figure \ref{fig:dpi1}). Depending on the input $X=x$, {c}hannel I emits the system $S$ in the state $\phi_x$, and {c}hannel II emits the system $S'$ in the state $\phi'_{x}={\mathcal E}_{S\rightarrow S'}(\phi_x)$. It is only necessary to verify that if a rate $R$ is achievable with $p(x)$ by {c}hannel II, $R$ is also achievable with $p(x)$ by {c}hannel I. Let $\{({\mathcal C}'^{(l)},{\mathcal D}'^{(l)})\}_l$ be a sequence of $(2^{lR},l)$ codes for {c}hannel II with the average error probability $P_e'^{(l)}$ and the tolerance $\tau'^{(l)}$. From the code $({\mathcal C}'^{(l)},{\mathcal D}'^{(l)})$, construct a $(2^{lR},l)$ code $({\mathcal C}^{(l)},{\mathcal D}^{(l)})$ for {c}hannel I by ${\mathcal C}^{(l)}={\mathcal C}'^{(l)}$ and ${\mathcal D}^{(l)}={\mathcal D}'^{(l)}\circ{\mathcal E}_{S\rightarrow S'}^{\otimes l}$. Here, ${\mathcal D}'^{(l)}\circ{\mathcal E}_{S\rightarrow S'}^{\otimes l}$ represents a process in which first ${\mathcal E}_{S\rightarrow S'}$ is applied to each of $S_1,\cdots,S_l$ individually and then the decoding measurement ${\mathcal D}'^{(l)}$ is performed on the total output system $S'_1\cdots S'_l$. The average error probability and the tolerance of this code are given by $P_e^{(l)}=P_e'^{(l)}$ and $\tau^{(l)}=\tau'^{(l)}$, respectively. Hence, if  $P_e'^{(l)}\rightarrow0$ and $\tau'^{(l)}\rightarrow0$, we also have $P_e^{(l)}\rightarrow0$ and $\tau^{(l)}\rightarrow0$, and thus $R$ is achievable with $p(x)$ by {c}hannel I.  \hfill{$\square$}
\label{prf:dpi}
\end{prf}

In general probabilistic theories, a measurement on a system $S$ without post-selection is described by a probabilistic map $\mathcal E_{\rm M}$ that maps states of $S$ into states of a classical system $T_S$. $T_S$ represents the register of the measurement outcomes. As a special case for Property \ref{prp:dpi}, we have $I_G(X:T_S)\leq I_G(X:S)$ under $\mathcal E_{\rm M}$, which is a generalization of Holevo's inequality. Let us define the accessible information $I_{\mathrm{acc}}(X:S)$ by
\begin{eqnarray}
I_{\mathrm{acc}}(X:S):=\max{I_C(X:T_S)}\;,
\end{eqnarray}
where the maximization is taken over all possible measurements on $S$. Then we have $0\leq I_{\mathrm{acc}}(X:S)\leq I_G(X:S)$.

\begin{figure}[t]
\centerline{\includegraphics[scale=0.44]{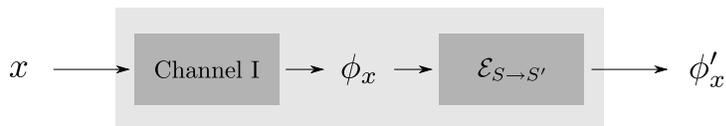}} 
\caption{{C}hannel II defined as the combination of {c}hannel I and ${\mathcal E}_{S\rightarrow S'}$.}
\label{fig:dpi1}
\end{figure}

To summarize, the generalized mutual information satisfies the following properties.
\begin{itemize}
\item {\it Symmetry}: $I_G(X:S)=I_G(S:X)$.
\item {\it Nonnegativity}: $I_G(X:S)\geq0$
\item {\it Consistency}: When $S$ is a classical system, $I_G(X:S)=I_C(X:S)$.
\item {\it Data Processing Inequality}: $I_G(X:S)\geq I_G(X':S')$ under local stochastic maps ${\mathcal E}_{X\rightarrow X'}$ and  ${\mathcal E}_{S\rightarrow S'}$ that contain no post-selection.
\end{itemize}
Thus, from Theorem \ref{thm:icsat} and Theorem \ref{thm:icvio}, we conclude that the chain rule of the generalized mutual information should be violated in any supernonlocal theory. Conversely, the chain rule implies Tsirelson's bound.

Throughout the rest of this paper, we use the generalized mutual information (GMI) given by Definition \ref{def:igdef}.

\section{Quantum mutual information}
\label{sec:quantum}

The quantum mutual information between a classical system $X$ and a quantum system $S$ is defined by
\begin{eqnarray}
I_Q(X:S)_{\hat \rho}:=H(S)_{\bar \rho}-\sum_{x\in{\mathcal X}}p(x)H(S)_{{\hat \rho}_x}\;,
\end{eqnarray}
where
\begin{eqnarray}
{\hat \rho}=\sum_{x\in{\mathcal X}}p(x)\proj{x}^X\otimes{\hat \rho}_x^S\;,\;\;\;\inpro{x}{x'}=\delta_{xx'}\;,\\
{\bar \rho}=\sum_{x\in{\mathcal X}}p(x){\hat \rho}_x\;,
\end{eqnarray}
and $H(S)$ is the von Newmann entropy. Note that, in quantum theory, a classical system is described by a Hilbert space in which we only consider a set of orthogonal pure states. With a slight generalization of the Holevo-Schumacher-Westmoreland theorem, it is shown that the GMI is a generalization of the quantum mutual information. 

\begin{thm}
In quantum theory, the GMI coincides with the quantum mutual information, i.e.,
\begin{eqnarray}
I_G(X:S)_{\Gamma_{\hat \rho}}=I_Q(X:S)_{\hat \rho}
\end{eqnarray}
where
\begin{eqnarray}
{\hat \rho}=\sum_{x\in{\mathcal X}}p(x)\proj{x}^{X}\otimes{\hat \rho}_x^S
\end{eqnarray}
and $\Gamma_{\hat \rho}=\{p(x),{\hat \rho}_x\}_{x\in{\mathcal X}}$\;.
\end{thm}

\begin{prf}
To prove this, it is only necessary to verify the following two statements:
\begin{enumerate}
\item A rate $R$ is achievable with $p(x)$ if $R<I_Q(X:S)_{\hat \rho}$,
\item A rate $R$ is achievable with $p(x)$ only if $R\leq I_Q(X:S)_{\hat \rho}$.
\end{enumerate}
The first statement is proved in \cite{schu1,schu2} by using random code generation, and the second statement is proved in the following way. Consider a $(2^{lR},l)$ code and suppose that Alice's message $W=1,\cdots,2^{lR}$ is uniformly distributed. Similarly to (\ref{eq:iexi1}), we have
\begin{eqnarray}
lR=H'(W)=I'(W:\hat{W})+H'(W|\hat{W})\leq I_Q'(X^l:S^l)+P_e^{(l)}lR+1\;.
\label{eq:igiq1}
\end{eqnarray}
Here, we use the data processing inequality. We also have
\begin{eqnarray}
I_Q'(X^l:S^l)&=H'(S^l)-H'(S^l|X^l)=H'(S^l)-\sum_{k=1}^lH'(S_k|X_k)\nonumber\\
		&\leq\sum_{k=1}^l(H'(S_k)-H'(S_k|X_k))=\sum_{k=1}^lI_Q'(X_k:S_k)\nonumber\\
		&=lI_Q'(X:S|K)=lI_Q'(X,K:S)-lI_Q'(K:S)\nonumber\\
		&\leq lI_Q'(X,K:S)=lI_Q'(X:S)\;.
\label{eq:igiq2}
\end{eqnarray}
In the first line, we use the fact that the state of $S_k$ depends only on $X_k$. The first inequality is from the subadditivity of the von Neumann entropy. The last equality holds since $K\rightarrow X\rightarrow S$ forms a Markov chain. From (\ref{eq:igiq1}) and (\ref{eq:igiq2}), we obtain
\begin{eqnarray}
P_e^{(l)}\geq1-\frac{I_Q'(X:S)}{R}-\frac{1}{lR}\;.
\end{eqnarray}
If $R$ is achievable with $p(x)$, there exists a sequence of $(2^{lR},l)$ codes satisfying $P_e^{(l)}\rightarrow0$ and $I_Q'(X:S)\rightarrow I_Q(X:S)_{\rho}$ when $l\rightarrow \infty$. Thus $R\leq I_Q(X:S)_{\rho}$.\hfill{$\square$}
\end{prf}

\section{No-supersignalling condition} 
\label{sec:cr}

In this section, to further investigate the derivation of Tsirelson's bound from information causality, we formulate a principle that we call the {{\it no-supersignalling condition}} by using the GMI. Suppose that Alice is trying to send to distant Bob information about $n$ independent classical bits $X_1,\cdots,X_n$, under the condition that they can only use a $m$ bit classical communication $\vec{M}$ from Alice to Bob and a supplementary resource of correlations shared in advance (see Figure \ref{nsssit}). The situation is similar to the setting of information causality described in Section \ref{sec:ic}, but now, we do not introduce random access coding. Instead, we evaluate Bob's information gain by $I_G(\vec{X}:\vec{M},B)$. We {say} that the no-supersignalling condition is satisfied if 
\begin{eqnarray}
I_G(\vec{X}:\vec{M},B)\leq m
\label{eq:nss}
\end{eqnarray}
holds for all $m\geq 0$. The condition indicates that {{\it the assistance of correlations cannot increase the capability of classical communication}}. It is a direct formulation of the {\it original} concept of information causality that ``$m$ bits of classical communication cannot produce more than $m$ bits of information gain''. In what follows, we prove that the no-supersignalling condition is equivalent to the no-signalling condition. It indicates that information causality and no-supersignalling are different.

\begin{figure}[t]
\centerline{\includegraphics[scale=0.55]{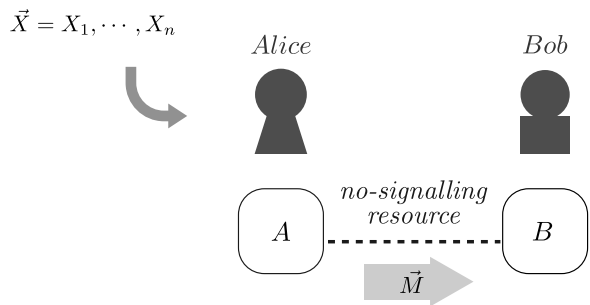}} 
\caption{The situation that the no-supersignalling condition refers to. The amount of information about $\vec{X}$ contained in $\vec{M}$ and $B$ is quantified by $I_G(\vec{X},\vec{M},B)$.}
\label{nsssit}
\end{figure}
 
\begin{lmm}
For any classical systems $X$, $Y$ and any general probabilistic system $S$, if $I_{\mathrm{acc}}(X:S)=0$ then $I_{\mathrm{acc}}(X:S,Y)\leq H(Y)$. 
\label{lmm:accy}
\end{lmm}

\begin{prf}
Consider a channel with an input system $X$ and two output systems $S$ and $Y$ (see Figure \ref{fig:channel2}). Let $\mathcal Z$ be the set of all measurements on $S$, and $p(t|x,y,z)$ be the probability of obtaining the outcome $t$ when the measurement $z\in\mathcal Z$ is performed on the system $S$ in the state $\phi_{xy}$. To achieve $I_{\mathrm{acc}}(X:S,Y)$, the receiver performs a measurement on $S$ possibly depending on $Y$. Let $z(y)$ be the optimal choice of the measurement when $Y=y$. The probability of obtaining the outcome $t$ when $X=x$ and $Y=y$ is given by
\begin{eqnarray}
p_1(t|x,y):=p(t|x,y,z(y))\;.
\end{eqnarray}
We define
\begin{eqnarray}
p_1(t,x,y):=p(x,y)p_1(t|x,y)=p(x,y)p(t|x,y,z(y))\;.
\end{eqnarray}
The condition $I_{\mathrm{acc}}(X:S)=0$ implies that for all $z\in\mathcal Z$,
\begin{eqnarray}
\sum_{y}p(x,y)p(t|x,y,z)=p(x)p_2(t|z)\;,
\end{eqnarray}
where
\begin{eqnarray}
p_2(t|z):=\sum_{x,y}p(x,y)p(t|x,y,z)\;.
\end{eqnarray}
Thus we obtain
\begin{eqnarray}
p_1(t,x,y)&=p(x,y)p(t|x,y,z(y))\nonumber\\
		&\leq \sum_{y'}p(x,y')p(t|x,y',z(y))\nonumber\\
		&=p(x)p_2(t|z(y))\;.
\label{eq:p1leq}
\end{eqnarray}
The accessible information $I_{\mathrm{acc}}(X:S,Y)$ is equal to the mutual information $I_C(X:T,Y)$ calculated for the probability distribution $p_1(t,x,y)$. Therefore
\begin{eqnarray}
I_{\mathrm{acc}}(X:S,Y)&=&I_C(X:T,Y)_{p_1}\nonumber\\
			&=&\sum_{t,x,y}p_1(t,x,y)\log{\frac{p_1(t,x,y)}{p(x)p_1(t,y)}}\nonumber\\
			&=&H(Y)+\sum_{t,x,y}p_1(t,x,y)\log{\frac{p_1(t,x,y)p(y)}{p(x)p_1(t,y)}}\nonumber\\
			&\leq&H(Y)+\sum_{t,x,y}p_1(t,x,y)\log{\frac{p(x)p(y)p_2(t|z(y))}{p(x)p_1(t,y)}}\nonumber\\
			&=&H(Y)-\sum_{t,y}p_1(t,y)\log{\frac{p_1(t,y)}{p_2(t,y)}}\nonumber\\
			&=&H(Y)-D(p_1(t,y)\|p_2(t,y))\nonumber\\
			&\leq&H(Y)\;.\nonumber
\end{eqnarray}
In the first inequality, we used (\ref{eq:p1leq}). In the next equality we defined a probability distribution $p_2(t,y):=p_2(t|z(y))p(y)$. The last inequality is from the nonnegativity of the relative entropy. \hfill{$\square$}
\end{prf}

\begin{figure}[t]
\centerline{\includegraphics[scale=0.44]{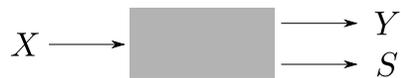}} 
\caption{The channel that we consider to prove Lemma \ref{lmm:accy}. For each pair of the input $X=x$ and the output $Y=y$, the corresponding state $\phi_{xy}$ of the output system $S$ is determined.}
\label{fig:channel2}
\end{figure}

\begin{thm}
The no-supersignalling condition defined in terms of the GMI (\ref{eq:nss}) is equivalent to the no-signalling condition.
\label{thm:noviolation}
\end{thm}
\begin{prf}
Consider a $(2^{lR},l)$ code for the channel presented in Figure \ref{fig:channel2} and let $X=\vec{X}$, $Y=\vec{M}$ and $S=B$. Suppose that Alice's message is uniformly distributed. By Fano's inequality, we have
\begin{eqnarray}
I'(W:\hat{W})\geq lR-1-P_e^{(l)}lR\;.
\end{eqnarray}
By the data processing inequality, we also have
\begin{eqnarray}
I'(W:\hat{W})\leq I'(X^l:Y^l,T_{S^l})\leq I'_{acc}(X^l:Y^l,S^l).
\end{eqnarray}
From the no-signalling condition, we have $I'_{acc}(X^l:S^l)=0$. From Lemma \ref{lmm:accy}, we obtain
\begin{eqnarray}
I'_{acc}(X^l:Y^l,S^l)\leq H'(Y^l)\;,
\end{eqnarray}
and thus
\begin{eqnarray}
I'(W:\hat{W})\leq H'(Y^l)\leq&lH'(Y)\;.
\end{eqnarray} 
Hence we obtain
\begin{eqnarray}
(1-P_e^{(l)})R\leq H'(Y)+\frac{1}{l}\;.
\end{eqnarray}
If $R$ is achievable with $p(x)$, there exists a sequence of $(2^{lR},l)$ codes that satisfies $P_e^{(l)}\rightarrow0$ and $H'(Y)\rightarrow H(Y)$ when $l\rightarrow \infty$. Thus, for any $R$ that is achievable with $p(x)$, we have $R\leq H(Y)$. It implies $I_G(X:Y,S)\leq H(Y)$ and thus $I_G(\vec{X}:\vec{M},B)\leq m$. Conversely, for $m=0$, the no-supersignalling condition $I_G(X:B)=0$ implies the no-signalling condition. \hfill{$\square$}
\end{prf}

\section{The difference between no-supersignalling and information causality}
\label{sec:nec}

In this section, we discuss the relation among information causality, no-supersignalling, Tsirelson's bound and the chain rule. Let us define
\begin{eqnarray}
\Delta_{\rm NSS}&:=&I_G(\vec{X}:\vec{M},B)-m\;,\\
\Delta_{\rm IC}&:=&J-m\;,\\
\Delta'&:=&\Delta_{\rm IC}-\Delta_{\rm NSS}=J-I_G(\vec{X}:\vec{M},B)\;.
\end{eqnarray}
$\Delta_{\rm NSS}$ quantifies how much the capability of classical communication is increased by the assistance of nonlocal correlations. {No-supersignalling is equivalent to $\Delta_{\rm NSS}\leq0$, and information causality is equivalent to $\Delta_{\rm IC}\leq0$}. $\Delta'$ quantifies the difference between no-supersignalling and information causality.
 
Theorem \ref{thm:icvio} states that, if Tsirelson's bound is violated, we have $\Delta_{\rm IC}>0$. Therefore violation of Tsirelson's bound implies at least either $\Delta_{\rm NSS}>0$ or $\Delta'>0$. Then which does violation of Tsirelson's bound imply, $\Delta_{\rm NSS}>0$ or $\Delta'>0$ ? As we proved in Section \ref{sec:cr}, $\Delta_{\rm NSS}\leq0$ is satisfied by all no-signalling theories. Thus violation of Tsirelson's bound only implies $\Delta'>0$. Therefore, {Tsirelson's bound is not derived from the condition that the assistance of nonlocal correlations does not increase the capability of classical communication. Instead, Tsirelson's bound is derived from the nonpositivity of $\Delta'$} (see Figure \ref{fig:icrelation}). Let us further define
\begin{eqnarray}
\Delta_{\rm CR}&:=&\sum_{k=1}^{n}I_G(X_k:\vec{M},B,X_1,\cdots,X_{k-1})-I_G(\vec{X}:\vec{M},B)\;.
\end{eqnarray}
The chain rule is equivalent to $\Delta_{\rm CR}=0$. By the data processing inequality, we always have $\Delta_{\rm CR}\geq\Delta'$. Thus the chain rule implies Tsirelson's bound\footnote{Another way to show this is to observe that the data processing inequality and the no-supersignalling condition imply $\Delta_{\rm CR}\geq\Delta_{\rm IC}$.} through imposing $\Delta'\leq\Delta_{\rm CR}=0$.

Let $X$ and $Y$ be two classical systems and $S$ be a general probabilistic system. The chain rule of the GMI is given by
\begin{eqnarray}
I_G(X,Y:S)+I_G(X:Y)=I_G(X:S)+I_G(Y:S,X)\;.
\label{eq:cr2}
\end{eqnarray}
Each term in (\ref{eq:cr2}) has an operational meaning as {an} information transmission rate by definition. The relation is satisfied in both classical and quantum theory, but is violated in all supernonlocal theories. Thus we can conclude that this highly nontrivial relation gives a strong restriction on the underlying physical theories. However, the operational meaning of this relation is not clear so far.

\begin{figure}[t]
 \begin{center}
  \includegraphics[scale=0.75]{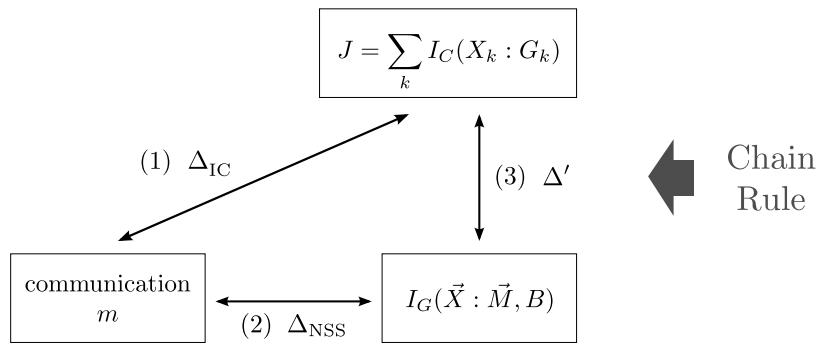}
  \caption{The relation between no-supersignalling and information causality, and the chain rule. Information causality refers to the gap in (1) represented by $\Delta_{\rm IC}$. No-supersignalling refers to the gap in (2) represented by $\Delta_{\rm NSS}$, and is irrelevant to Tsirelson's bound. The gap in (3) represented by $\Delta'$ is crucial in the derivation of Tsirelson's bound. $\Delta'$ is bounded above by zero if the chain rule is satisfied.}
  \label{fig:icrelation}
 \end{center}
\end{figure}

\section{Restriction on one gbit state space}
\label{sec:ex}
To investigate how the chain rule of the GMI imposes a restriction on {p}hysical theories, we consider a gbit {--} the counterpart of a qubit in general probabilistic theories \cite{gbit}. Here, we {do not make assumptions about} a gbit such as the dimension of the state space, {or} the possibility or impossibility of various measurements and transformations. Instead, we define a gbit as the minimum unit of information in the theory, and require that the classical information capacity of one gbit is not more than one bit. Thus we require
\begin{eqnarray}
I_G(X:S_{\rm 1gb})\leq1
\label{eq:gbcap}
\end{eqnarray}
for any classical system $X$. When $X$ is a classical system composed of two independent and uniformly random bits $X_0$ and $X_1$, we have
\begin{eqnarray}
I_G(X_0,X_1:S_{\rm 1gb})\leq1\;.
\end{eqnarray}
By the chain rule, we have
\begin{eqnarray}
I_G(X_0,X_1:S_{\rm 1gb})=I_G(X_0:S_{\rm 1gb})+I_G(X_1:S_{\rm 1gb},X_0)\;.
\end{eqnarray}
By the data processing inequality, we also have
\begin{eqnarray}
I_G(X_0:S_{\rm 1gb})+I_G(X_1:S_{\rm 1gb},X_0)\geq I_{\mathrm{acc}}(X_0:S_{\rm 1gb})+I_{\mathrm{acc}}(X_1:S_{\rm 1gb})\;.
\end{eqnarray}
Thus the chain rule implies
\begin{eqnarray}
I_{\mathrm{acc}}(X_0:S_{\rm 1gb})+I_{\mathrm{acc}}(X_1:S_{\rm 1gb})\leq1\;.
\label{eq:1gb}
\end{eqnarray}
We consider success probabilities of the decoding measurements on $S_{\rm 1gb}$ for $X_0$ and $X_1$. For simplicity, we assume  that the optimal measurement performed on $S_{\rm 1gb}$ to decode $X_0$ or $X_1$ has two outcomes $t=0,1$. Let $P(t|m,x_0,x_1)$ be the probability of obtaining the outcome $t$ when $X_0=x_0$, $X_1=x_1$ and the measurement $m$ is performed. The index $m=0,1$ corresponds to the optimal measurement for decoding $X_0$, $X_1$, respectively. The list of all probabilities $\{P(t|m,x_0,x_1)\}_{t,m,x_0,x_1=0,1}$ can be regarded as representing a ``state''. We compare the state space of a qubit and the state space determined by (\ref{eq:1gb}). For further simplicity, we assume that for all $x_0$ and $x_1$,
\begin{eqnarray}
P(t=x_0|m=0,x_0,x_1)=\frac{1+\alpha}{2}\;\;\;\;(0\leq\alpha\leq1)\;,\nonumber\\
P(t=x_1|m=1,x_0,x_1)=\frac{1+\beta}{2}\;\;\;\;(0\leq\beta\leq1)\;.\:\nonumber
\end{eqnarray}
Then we have
\begin{eqnarray}
I_{\mathrm{acc}}(X_0:S_{\rm 1gb})&=I_C(x_1:t|m=0)=1-H(x_0|t,m=0)\nonumber\\
&=1-H(x_0\oplus t|m=0)=1-h\left(\frac{1+\alpha}{2}\right)\;,
\label{eq:1gb1}
\end{eqnarray}
and
\begin{eqnarray}
I_{\mathrm{acc}}(X_1:S_{\rm 1gb})=1-h\left(\frac{1+\beta}{2}\right)\;.
\label{eq:1gb2}
\end{eqnarray}
Here, $h(x)$ is the binary entropy defined by $h(x):=-x\log{x}-(1-x)\log{(1-x)}$. From (\ref{eq:1gb}), (\ref{eq:1gb1}) and (\ref{eq:1gb2}), we have
\begin{eqnarray}
h\left(\frac{1+\alpha}{2}\right)+h\left(\frac{1+\beta}{2}\right)\geq1\;.
\label{eq:1gbf}
\end{eqnarray}
This inequality gives a restriction on the state space of one gbit (see Figure \ref{fig:gbitbound}). It is shown in Appendix B that in the case of one qubit, the obtainable region is given by $\alpha^2+\beta^2\leq1$.

\begin{figure}[t]
 \begin{center}
  \includegraphics[scale=0.5]{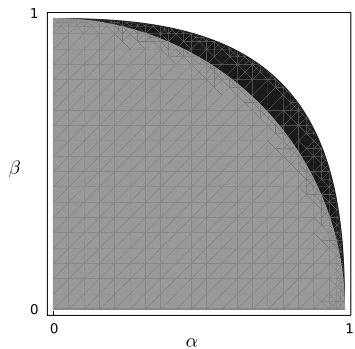}
  \caption{Comparison of the state space of a qubit and the boundary given by the chain rule. The grey region indicates the state space of a qubit given by $\alpha^2+\beta^2\leq1$. The black region in addition to the grey region indicates the region defined by (\ref{eq:1gbf}).}
  \label{fig:gbitbound}
 \end{center}
\end{figure}

\section{Conclusions and discussions}
\label{sec:conclusion}
We {have} defined a generalized mutual information (GMI) between a classical system and a general probabilistic system. Since the definition is based on the channel coding theorem, the GMI inherently has an operational meaning as {an} information transmission rate. We showed that the GMI coincides with the quantum mutual information if the {output} system is quantum. The GMI satisfies nonnegativity, symmetry, the data processing inequality, and the consistency with the classical mutual information, but does not necessarily satisfy the chain rule.

{U}sing the GMI, we have analyzed the derivation of Tsirelson's bound from information causality defined in terms of the efficiency of nonlocality-assisted random access coding. We showed that the chain rule of the GMI, which is satisfied in both classical and quantum theory, is violated in any theory in which the existence of nonlocal correlations exceeding Tsirelson's bound is allowed. Thus we conclude that the chain rule of the GMI implies Tsirelson's bound. 

We formulated a condition, the no-supersignalling condition, which states that the assistance of nonlocal correlations does not increase the capability of classical communication. We proved that this condition is equivalent to the no-signalling condition. We also clarified the relation among no-supersignalling, information causality, Tsirelson's bound and the chain rule.

The derivation of Tsirelson's bound from information causality proposed in \cite{ic1} is remarkable in that the Tsirelson's bound is exactly derived and that {to do so} we only need {the five properties of the mutual information}. However, information causality is different from the condition that ``$m$ bits of classical communication cannot produce more than $m$ bits of information gain''. This derivation shows that several laws of Shannon theory{\footnote{{By Shannon theory we mean the theoretical framework composed of various theorems on the asymptotic coding rate of the sources and the channels.}}}, represented by the five properties of the mutual information, taken together impose a strong restriction on the underlying physical theory. If we take the GMI as {the} definition of the mutual information, it reduces to the statement that ``a law of Shannon theory, namely the chain rule of the GMI, imposes a strong restriction on the underlying physical theory''.

Although the operational meaning of the GMI is clear, we have not yet succeeded in finding {a} clear operational meaning of the chain rule. In classical and quantum Shannon theory, the chain rule appears in a lot of proofs of coding theorems. Therefore, investigation of the meaning of the chain rule would lead us to a {better} understanding of the informational foundations of quantum mechanics.  {O}n the other hand, our definition of the generalized mutual information {is not} the only way to generalize the quantum mutual information. It would also be fruitful to seek {out} other operationally motivated definitions of the generalized mutual information and compare them{.}

\ack
We thank Takanori Sugiyama and Salman Beigi for useful discussions. We also thank referees for useful comments. This work was supported by Project for Developing Innovation Systems of Ministry of Education, Culture, Sports, Science and Technology (MEXT), Japan. MM acknowledges support from JSPS by KAKENHI (Grant No. 23540463).

\appendix

\section{Data processing inequality}
\label{sec:dpi}
We prove the latter part of Theorem \ref{prp:dpi}, which states that under any local stochastic map ${\mathcal E}_{X\rightarrow X'}$ that contains no post-selection, we have
\begin{eqnarray}
I_G(X:S)\geq I_G(X':S)\;.
\label{eq:dpix}
\end{eqnarray}
The effect of ${\mathcal E}_{X\rightarrow X'}$ is determined by a conditional probability distribution $p_{\mathcal E}(x'|x)$, where $x$ and $x'$ denote the states of $X$ and $X'$, respectively. Let $\{p(x),\phi_x\}_{x\in{\mathcal X}}$ be the state of $XS$ before applying ${\mathcal E}_{X\rightarrow X'}$. We can define probability distributions $p_{\mathcal E}(x,x')=p(x)p_{\mathcal E}(x'|x)$,  $p(x')=\sum_{x}p_{\mathcal E}(x,x')$ and $p_{\mathcal E}(x|x')=p_{\mathcal E}(x,x')/p(x')$ for $x\in{\mathcal X}$ and $x'\in{\mathcal X}'$. The state of $X'S$ after applying ${\mathcal E}_{X\rightarrow X'}$ is $\{p(x'),\phi_{x'}\}_{x'\in{\mathcal X}'}$, where $\phi_{x'}$ is the mixture of $\phi_x$ with the probability given by $p_{\mathcal E}(x|x')$. We assume that $|{\mathcal X}|,|{\mathcal X}'|<\infty$. 

To prove (\ref{eq:dpix}), consider two channels, {c}hannel I and {c}hannel III (see Figure \ref{fig:dpi2}). {C}hannel I outputs the system $S$ in the state $\phi_x$ according to the input $X=x$, and {c}hannel III outputs the system $S$ in the state $\phi_{x'}$ according to the input $X'=x'$. It is only necessary to show that if a rate $R$ is achievable with $p(x')$ by {c}hannel III, $R$ is also achievable with $p(x)$ by {c}hannel I. Consider a sequence of $(2^{lR},l)$ codes $({\mathcal C}'^{(l)},{\mathcal D}'^{(l)})$ for {c}hannel III that satisfies
\begin{enumerate}
\item $P_e'^{(l)}\rightarrow0$ when $l\rightarrow\infty$,
\item $\tau'^{(l)}\rightarrow 0$ when $l\rightarrow\infty$.
\end{enumerate}
Such a sequence exists if $R$ is achievable with $p(x')$ by {c}hannel III. From the code $({\mathcal C}'^{(l)},{\mathcal D}'^{(l)})$, we randomly construct  $(2^{lR},l)$ codes $({\mathcal C}^{(l)},{\mathcal D}^{(l)})$ for {c}hannel I in the following way. 
\begin{itemize}
\item For any $w$ and $k$ $(1\leq w\leq 2^{lR},1\leq k\leq l)$, generate the codeletter $x_k(w)$ randomly and independently according to the probability distribution $P(x_k(w)=x)=p_{\mathcal E}(x|x'_k(w))$.
\item Regardless of the randomly generated codebook ${\mathcal C}^{(l)}$, use the same decoding measurement ${\mathcal D}^{(l)}={\mathcal D}'^{(l)}$.
\end{itemize}
Let $P_e^{{\mathcal C}^{(l)}}$ be the average error probability of the code $({\mathcal C}^{(l)},{\mathcal D}^{(l)})$ defined by
\begin{eqnarray}
P_e^{{\mathcal C}^{(l)}}:=\frac{1}{2^{lR}}\sum_{u=1}^{2^{lR}}P({\hat W}\neq u|W=u,{\mathcal C}^{(l)})\;.
\end{eqnarray}
Averaging $P_e^{{\mathcal C}^{(l)}}$ over all codebooks ${\mathcal C}^{(l)}$ that are randomly generated, we obtain
\begin{eqnarray}
{\bar P}_e^{(l)}:=\sum_{{\mathcal C}^{(l)}}P({\mathcal C}^{(l)})\:P_e^{{\mathcal C}^{(l)}}\;,
\label{eq:defpel}
\end{eqnarray}
where $P({\mathcal C}^{(l)})$ is the probability of obtaining the codebook ${\mathcal C}^{(l)}$ as a result of random code generation. In Lemma A.1, we show that ${\bar P}_e^{(l)}\rightarrow0$ in the limit of $l\rightarrow\infty$. In Lemma A.2, we prove that for sufficiently large $l$, the tolerance $\tau^{(l)}$ of the codebook ${\mathcal C}^{(l)}$ is almost equal to 0 with arbitrarily high probability. Finally, we give the proof for (\ref{eq:dpix}) in Theorem A.3.

\begin{figure}[t]
\centerline{\includegraphics[scale=0.44]{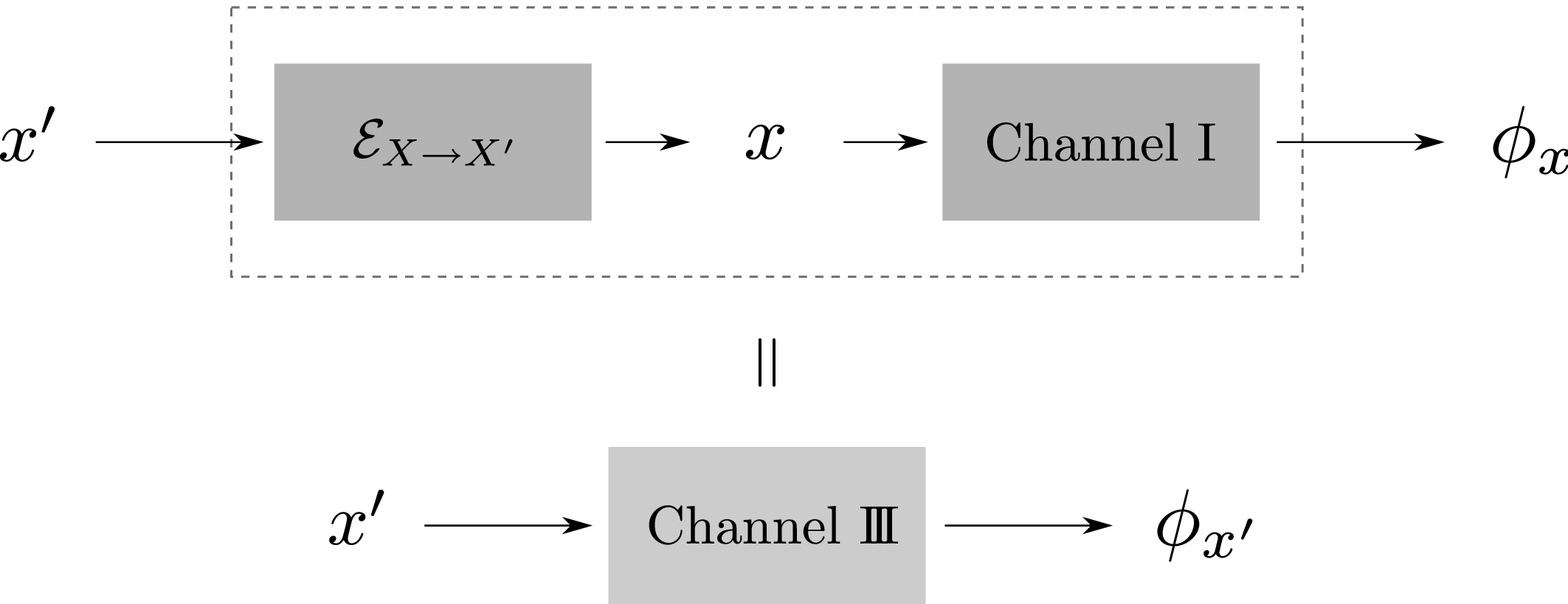}} 
\caption{{C}hannel III defined as the combination of ${\mathcal E}_{X\rightarrow X'}$ and {c}hannel I. This channel as a whole is equivalent to a channel with the input $x'$ and the output $\phi_{x'}$.}
\label{fig:dpi2}
\end{figure}

\begin{lmmA1}
\begin{eqnarray}
\lim_{l\rightarrow\infty}{\bar P}_e^{(l)}=0\;.
\end{eqnarray}
\label{lmm:pel0}
\end{lmmA1}

\begin{prf}
${\bar P}_e^{(l)}$ defined by (\ref{eq:defpel}) is calculated to 
\begin{eqnarray}
{\bar P}_e^{(l)}&=&\sum_{{\mathcal C}^{(l)}}P({\mathcal C}^{(l)})\times\frac{1}{2^{lR}}\sum_{u=1}^{2^{lR}}P({\hat W}\neq u|W=u,{\mathcal C}^{(l)})\nonumber\\
		&=&\frac{1}{2^{lR}}\sum_{u=1}^{2^{lR}}\sum_{{\mathcal C}^{(l)}}P({\mathcal C}^{(l)})P({\hat W}\neq u|W=u,{\mathcal C}^{(l)})\nonumber\\
		&=&\frac{1}{2^{lR}}\sum_{u=1}^{2^{lR}}{\bar P}({\hat W}\neq u|W=u)\;,
\end{eqnarray}
where
\begin{eqnarray}
{\bar P}({\hat W}\neq u|W=u):=\sum_{{\mathcal C}^{(l)}}P({\mathcal C}^{(l)})P({\hat W}\neq u|W=u,{{\mathcal C}^{(l)}})\;.
\end{eqnarray}
The codebook ${\mathcal C}^{(l)}$ is determined by the codeletters $x_k(w)$ $(1\leq w\leq 2^{lR},1\leq k\leq l)$. Due to the way of randomly generating the code, the probability of obtaining the codebook ${\mathcal C}^{(l)}$ such that $x_k(w)=\xi_{wk}$ $(1\leq w\leq 2^{lR},1\leq k\leq l)$ is given by
\begin{eqnarray}
P({\mathcal C}^{(l)})&=&P(\{x_k(w)\}_{w,k}=\{\xi_{wk}\}_{w,k})\nonumber\\
				&=&\prod_{w=1}^{2^{lR}}\prod_{k=1}^{l}P(x_k(w)=\xi_{wk})\nonumber\\
				&=&\prod_{w=1}^{2^{lR}}\prod_{k=1}^{l}p_{\mathcal E}(x=\xi_{wk}|x'=x'_k(w))\;.
\end{eqnarray} 
Let $D(\phi_{x_1}\cdots\phi_{x_l})$ be the result of the decoding measurement ${\mathcal D}^{(l)}$ on the composite system $S_1\cdots S_l$ in the state $\phi_{x_1}\cdots\phi_{x_l}$. We have
\begin{eqnarray}
\fl
P({\hat W}\neq u|W=u,{\mathcal C}^{(l)})&=P(D(\phi_{x_1(u)}\cdots\phi_{x_l(u)})\neq u|\{x_k(w)\}_{w,k}=\{\xi_{wk}\}_{w,k})\nonumber\\
&=P(D(\phi_{\xi_{u1}}\cdots\phi_{\xi_{ul}})\neq u)\;,
\end{eqnarray}
and we obtain
\begin{eqnarray}
\fl{\bar P}({\hat W}\neq u|W=u)\nonumber\\
\fl=\sum_{\{\xi_{wk}\}_{w,k}}P(D(\phi_{x_1(u)}\cdots\phi_{x_l(u)})\neq u|\{x_k(w)\}_{w,k}=\{\xi_{wk}\}_{w,k})\times P(\{x_k(w)\}_{w,k}=\{\xi_{wk}\}_{w,k})\nonumber\\
\fl=\sum_{\{\xi_{uk}\}_{k}}P(D(\phi_{\xi_{u1}}\cdots\phi_{\xi_{ul}})\neq u)\times P(\{x_k(u)\}_{k}=\{\xi_{uk}\}_{k})\nonumber\\
\fl=\sum_{\{\xi_{uk}\}_{k}}P(D(\phi_{\xi_{u1}}\cdots\phi_{\xi_{ul}})\neq u)\times \prod_{k=1}^{l}p_{\mathcal E}(x=\xi_{uk}|x'=x'_k(u))\;.
\label{eq:pecompare1}
\end{eqnarray}
On the other hand, the error probability for the message $w$ when {c}hannel III is used with the code $({\mathcal C}'^{(l)},{\mathcal D}'^{(l)})$ is given by
\begin{eqnarray}
&&P'({\hat W}\neq u|W=u)\nonumber\\
&=&P(D(\phi_{x'_1(u)}\cdots\phi_{x'_l(u)})\neq u)\nonumber\\
&=&\sum_{\{x_{k}\}_{k}} P(D(\phi_{x_1}\cdots\phi_{x_l})\neq u)\times\prod_{k=1}^{l}p_{\mathcal E}(x=x_{k}|x'=x'_k(u))\;.
\label{eq:pecompare2}
\end{eqnarray}
From (\ref{eq:pecompare1}) and (\ref{eq:pecompare2}), we obtain
\begin{eqnarray}
{\bar P}({\hat W}\neq u|W=u)=P'({\hat W}\neq u|W=u)\;,
\end{eqnarray}
and consequently
\begin{eqnarray}
{\bar P}^{(l)}_{e}=P'^{(l)}_e\;.
\end{eqnarray}
Therefore ${\bar P}_e^{(l)}\rightarrow0$ when $l\rightarrow\infty$.\hfill{$\square$}
\end{prf}

\begin{lmmA2}
$\tau^{(l)}\rightarrow0$ in probability in the limit of $l\rightarrow \infty$.
\label{lmm:tau0}
\end{lmmA2}

\begin{prf}
Let $f(x)^{(l)}$ and $f(x')^{(l)}$ be the letter frequency of the codebook ${\mathcal C}^{(l)}$ and ${\mathcal C}'^{(l)}$, respectively. We have
\begin{eqnarray}
\fl\left|f(x)^{(l)}-p(x)\right|=\left|f(x)^{(l)}-\sum_{x'\in{\mathcal X}'}p_{\mathcal E}(x|x')p(x')\right|\nonumber\\
\fl\leq\left|f(x)^{(l)}-\sum_{x'\in{\mathcal X}'}f(x')^{(l)}p_{\mathcal E}(x|x')\right|+\left|\sum_{x'\in{\mathcal X}'}f(x')^{(l)}p_{\mathcal E}(x|x')-\sum_{x'\in{\mathcal X}'}p_{\mathcal E}(x|x')p(x')\right|\nonumber\\
\fl\leq\left|f(x)^{(l)}-\sum_{x'\in{\mathcal X}'}f(x')^{(l)}p_{\mathcal E}(x|x')\right|+\sum_{x'\in{\mathcal X}'}p_{\mathcal E}(x|x')\left|f(x')^{(l)}-p(x')\right|\nonumber\;.
\end{eqnarray}
Define
\begin{eqnarray}
f(x,x')^{(l)}:=\frac{|\{(k,w)|x_k(w)=x, x'_k(w)=x',1\leq k\leq l, 1\leq w\leq 2^{lR}\}|}{l\cdot2^{lR}}\nonumber
\end{eqnarray}
for $x\in{\mathcal X},x'\in{\mathcal X}'$.
By using the relation
\begin{eqnarray}
f(x)^{(l)}=\sum_{x'\in{\mathcal X}'}f(x')^{(l)}\frac{f(x,x')^{(l)}}{f(x')^{(l)}}\;,
\end{eqnarray}
we obtain
\begin{eqnarray}
\Delta(x)^{(l)}&:=&\left|f(x)^{(l)}-\sum_{x'\in{\mathcal X}'}f(x')^{(l)}p_{\mathcal E}(x|x')\right|\nonumber\\
&\leq&\sum_{x'\in{\mathcal X}'}f(x')^{(l)}\left|\frac{f(x,x')^{(l)}}{f(x')^{(l)}}-p_{\mathcal E}(x|x')\right|\;.
\end{eqnarray}
Applying the weak law of large numbers for each term in the sum, we have $\Delta(x)^{(l)}\rightarrow0\;(l\rightarrow\infty)$ in probability. We also have
\begin{eqnarray}
\sum_{x'\in{\mathcal X}'}p_{\mathcal E}(x|x')\left|f(x')^{(l)}-p(x')\right|\leq\:\tau'^{(l)}\cdot|{\mathcal X}'|
\end{eqnarray}
and thus
\begin{eqnarray}
\lim_{l\rightarrow\infty}\;\;\sum_{x'\in{\mathcal X}'}p_{\mathcal E}(x|x')\left|f(x')^{(l)}-p(x')\right|=0\;.
\end{eqnarray}
Therefore we obtain
\begin{eqnarray}
\tau^{(l)}=\max_x\left|f(x)^{(l)}-p(x)\right|\rightarrow0\qquad\textrm{\it in probability}\;.
\end{eqnarray}
\hfill{$\square$}
\end{prf}

\begin{thmA3}
$R$ is achievable with $p(x)$ by {c}hannel I.
\label{thm:dpix}
\end{thmA3}

\begin{prf}
Take arbitrary $\epsilon,\delta,\eta>0$. From Lemma A.1 and Lemma A.2, for sufficiently large $l$ we have
\begin{eqnarray}
{\bar P}_e^{(l)}<\epsilon
\end{eqnarray}
and
\begin{eqnarray}
{\rm Pr}\{\tau^{(l)}<\delta\}>1-\eta\;.
\end{eqnarray} 
Define $C^{(l)}_{\delta}:=\{{\mathcal C}^{(l)}|\tau^{(l)}<\delta\}$. The average error probability averaged over  all codebooks in $C^{(l)}_{\delta}$ is calculated to
\begin{eqnarray}
\fl \frac{\sum_{{\mathcal C}^{(l)}\in C^{(l)}_{\delta}}P({\mathcal C}^{(l)})P_e^{{\mathcal C}^{(l)}}}{\sum_{{\mathcal C}^{(l)}\in C^{(l)}_{\delta}}P({\mathcal C}^{(l)})}
=\frac{{\bar P}_e^{(l)}-\sum_{{\mathcal C}^{(l)}\notin C^{(l)}_{\delta}}P({\mathcal C}^{(l)})P_e^{{\mathcal C}^{(l)}}}{\sum_{{\mathcal C}^{(l)}\in C^{(l)}_{\delta}}P({\mathcal C}^{(l)})}
\leq\frac{{\bar P}_e^{(l)}}{\sum_{{\mathcal C}^{(l)}\in C^{(l)}_{\delta}}P({\mathcal C}^{(l)})}
<\frac{\epsilon}{1-\eta}\;.\nonumber
\end{eqnarray}
Thus there exists at least one codebook ${\mathcal C}^{(l)}\in C^{(l)}_{\delta}$ such that $P_e^{{\mathcal C}^{(l)}}<\epsilon'=\epsilon/(1-\eta)$ and, by definition,  $\tau^{(l)}<\delta$. Hence there exists a sequence of $(2^{lR},l)$ codes for {c}hannel I such that $P_e^{(l)}\rightarrow0$ and $\tau'^{(l)}\rightarrow 0$ when $l\rightarrow\infty$, and thus $R$ is achievable with $p(x)$ by {c}hannel I.\hfill{$\square$}
\end{prf}

\section{Beyond the global state assumption}
\label{sec:gsa}
In this appendix we generalize the results presented in the main sections to general probabilistic theories which do not satisfy the global state assumption. Suppose that there are $l$ independent copies of a channel that outputs the system $S$ in the state $\phi_x$ according to the input $X=x$. If the input sequence is $x_1\cdots x_l$, the state of the output system $S_1\cdots S_l$ is $\phi_{x_1}\cdots\phi_{x_l}$. However, without the global state assumption, this does not specify the ``global'' state of the composite system: it only specifies the state of the composite system for product measurements. Thus it is not sufficient to determine the rate of the channel. To avoid this difficulty, we introduce the notion of ``consistency'' of the states. Let $\Phi_{x_1\cdots x_l}$ be  a global state of $S_1\cdots S_l$. We say {\it $\Phi_{x_1\cdots x_l}$ is consistent with $\phi_{x_1}\cdots\phi_{x_l}$} if the two states exhibit the same statistics for any product measurement. $\Phi^{(l)}:=\{\Phi_{x_1\cdots x_l}\}_{x_1\cdots x_l\in{\mathcal X}^l}$ is said to be consistent with $\{\phi_{x_1}\cdots\phi_{x_l}\}_{x_1\cdots x_l\in{\mathcal X}^l}$ if $\Phi_{x_1\cdots x_l}$ is consistent with $\phi_{x_1}\cdots\phi_{x_l}$ for all $x_1\cdots x_l\in{\mathcal X}^l$. With a slight abuse of terminology, we say $\Phi:=\{\Phi^{(l)}\}_{l=1}^\infty$ is consistent with $\{\phi_x\}_{x\in{\mathcal X}}$ if $\Phi^{(l)}$ is consistent with $\{\phi_{x_1}\cdots\phi_{x_l}\}_{x_1\cdots x_l\in{\mathcal X}^l}$ for all $l$. Let $\Gamma_{\Phi}:=\{\Gamma_{\Phi}^{(l)}\}_{l=1}^\infty$ be the sequence of the channel $\Gamma_{\Phi}^{(l)}$ that outputs the system $S_1\cdots S_l$ in the state $\Phi_{x_1\cdots x_l}\in\Phi^{(l)}\in\Phi$ according to the input $X_1\cdots X_l=x_1\cdots x_l$.

\begin{dfnB1}
A rate $R$ is said to be achievable with $p(x)$ for $\Phi$ if there exists a sequence of $(2^{lR},l)$ codes $({\mathcal C}^{(l)},{\mathcal D}^{(l)})$ for $\Gamma_{\Phi}^{(l)}\in\Gamma_{\Phi}$ such that 
\begin{enumerate}
\item $P_e^{(l)}\rightarrow0$ when $l\rightarrow\infty$,
\item $\tau^{(l)}\rightarrow 0$ when $l\rightarrow\infty$.
\end{enumerate}
\end{dfnB1}

\begin{dfnB2}
A rate $R$ is said to be achievable with $p(x)$ if $R$ is achievable with $p(x)$ for all $\Phi$ that is consistent with $\{\phi_x\}_{x\in{\mathcal X}}$.
\end{dfnB2}

We define the generalized mutual information by Definition \ref{def:igdef} and its existence is proved by Theorem \ref{thm:existence}. The data processing inequality (Property \ref{prp:dpi}) is proved as follows.

\begin{prf}
The inequality $I_G(X:S)\geq I_G(X:S')$ under local transformation ${\mathcal E}_{S\rightarrow S'}$ is proved as follows.
\begin{eqnarray}
\fl I_G(X:S')\nonumber\\
\fl=\sup\{R|R\textrm{ is achievable for all }\Phi'\textrm{ that is consistent with }\{{\mathcal E}(\phi_{x})\}_{x\in{\mathcal X}}\}\nonumber\\
\fl\leq\sup\{R|R\textrm{ is achievable for }{\mathcal E}(\Phi)\textrm{ for all }\Phi\textrm{ that is consistent with }\{\phi_{x}\}_{x\in{\mathcal X}}\}\nonumber\\
\fl\leq\sup\{R|R\textrm{ is achievable for all }\Phi\textrm{ that is consistent with }\{\phi_{x}\}_{x\in{\mathcal X}}\}\nonumber\\
\fl=I_G(X:S)\;.
\end{eqnarray}
Here, ${\mathcal E}(\Phi):=\{{\mathcal E}^{\otimes l}(\Phi^{(l)})\}_{l=1}^\infty$ and ${\mathcal E}^{\otimes l}(\Phi^{(l)}):=\{{\mathcal E}^{\otimes l}(\Phi_{x_1\cdots x_l})\}_{x_1\cdots x_l\in{\mathcal X}^l}$. The first inequality comes from the fact that ${\mathcal E}(\Phi)$ is consistent with $\{{\mathcal E}(\phi_{x})\}_{x\in{\mathcal X}}$ if $\Phi$ is consistent with $\{\phi_{x}\}_{x\in{\mathcal X}}$. The second inequality is proved in the same way as the proof {presented} in page \pageref{prf:dpi}.

The inequality $I_G(X:S)\geq I_G(X':S)$ under local transformation ${\mathcal E}_{X\rightarrow X'}$ is proved as follows.
\begin{eqnarray}
\fl I_G(X':S)\nonumber\\
\fl=\sup\{R|R\textrm{ is achievable for all }\Phi'\textrm{ that is consistent with }\{\phi_{x'}\}_{x'\in{\mathcal X'}}\}\nonumber\\
\fl\leq\sup\{R|R\textrm{ is achievable for }\Phi_{X'}\textrm{ for all }\Phi\textrm{ that is consistent with }\{\phi_{x}\}_{x\in{\mathcal X}}\}\nonumber\\
\fl\leq\sup\{R|R\textrm{ is achievable for all }\Phi\textrm{ that is consistent with }\{\phi_{x}\}_{x\in{\mathcal X}}\}\nonumber\\
\fl=I_G(X:S)
\end{eqnarray}
Here, $\Phi_{X'}:=\{\Phi^{(l)}_{X'}\}_{l=1}^\infty$ and $\Phi^{(l)}_{X'}:=\{\Phi_{x'_1\cdots x'_l}\}_{x'_1\cdots x'_l\in{\mathcal X}'^l}$, where $\Phi_{x'_1\cdots x'_l}$ is the mixture of $\Phi_{x_1\cdots x_l}\in\Phi^{(l)}\in\Phi$ with the probability $\prod_{k=1}^lp_{\mathcal E}(x_k|x'_k)$. The first inequality comes from the fact that $\Phi_{X'}$ is consistent with $\{\phi_{x'}\}_{x'\in{\mathcal X'}}$ if $\Phi$ is consistent with $\{\phi_{x}\}_{x\in{\mathcal X}}$. The second inequality is proved in the same way as the proof in \ref{sec:dpi}, where $\phi_{x_1}\cdots\phi_{x_l}$ is replaced by $\Phi_{x_1\cdots x_l}$.
\hfill{$\square$}
\end{prf}

The equivalence of no-supersignalling and no-signalling (Theorem \ref{thm:noviolation}) is proved as follows.

\begin{prf}
Due to the no-signalling condition, there exists $\Phi$ that is consistent with $\{\phi_{xy}\}_{x\in{\mathcal X},y\in{\mathcal Y}}$, and satisfies $I'_{acc}(X^l:S^l)=0$ for all  $\Gamma_{\Phi}^{(l)}\in\Gamma_{\Phi}$. Here, $\Gamma_{\Phi}^{(l)}$ is a channel with an input system $X^l$ and two output systems $Y^l$ and $S^l$. According to the input $X^l=x^l$, the channel outputs $Y^l=y^l$ with the probability $\prod_{k=1}^lp(y_k|x_k)$ and the system $S^l$ in the state $\Phi_{x_1y_1\cdots x_ly_l}\in\Phi^{(l)}\in\Phi$. Consider a $(2^{lR},l)$ code for the channel. In the same way as the proof of Theorem \ref{thm:noviolation}, we have $(1-P_e^{(l)})R\leq H'(Y)+1/l$. If $R$ is achievable with $p(x)$ for $\Phi$, there exists a sequence of $(2^{lR},l)$ code for $\Gamma_{\Phi}^{(l)}$ that satisfies $P_e^{(l)}\rightarrow0$ and $H'(Y)\rightarrow H(Y)$ when $l\rightarrow \infty$. Thus, for any $R$ that is achievable with $p(x)$, we have $R\leq H(Y)$. It implies $I_G(X:Y,S)\leq H(Y)$ and thus $I_G(\vec{X}:\vec{M},B)\leq m$. Conversely, for $m=0$, the no-supersignalling condition $I_G(X:B)=0$ implies the no-signalling condition. \hfill{$\square$}
\end{prf}

\section{State space of a qubit}
Suppose that two independent and uniformly random bits $X_0, X_1$ are encoded into the state of a qubit ${\hat \rho}_{x_0x_1}$. Let $\{{\hat M}^m_t\}_{t=0,1}$ be the optimal measurement for decoding $X_m$ ($m=0,1$),
 where the mutual information $I_C(X_m:T)$ between $X_m$ and the measurement outcome $T$ is maximized when the measurement $m$ is performed. We assume that for all $x_0$ and $x_1$,
\begin{eqnarray}
P(t=x_0|m=0,x_0,x_1)={\rm tr}[{\hat M}^0_{x_0}{\hat \rho}_{x_0x_1}]=\frac{1+\alpha}{2}\;\;\;\,(0\leq\alpha\leq1)\;,\label{eq:bloch1}\\
P(t=x_1|m=1,x_0,x_1)={\rm tr}[{\hat M}^1_{x_1}{\hat \rho}_{x_0x_1}]=\frac{1+\beta}{2}\;\;\;\;(0\leq\beta\leq1)\;.\label{eq:bloch2}
\end{eqnarray}
In what follows, we prove that such a set of density operators $\{{\hat \rho}_{x_0x_1}\}_{x_0,x_1=0,1}$ and POVM operators $\{{\hat M}^m_t\}_{m,t=0,1}$ exists if and only if $\alpha^2+\beta^2\leq1$. Considering the parametrization of a qubit state using the Bloch sphere, the ``if'' part is obviously verified. The ``only if'' part is proved as follows. Let ${\bm r}_{x_0x_1}$ be the Bloch vector representation of ${\hat \rho}_{x_0x_1}$ and ${\bm u}$, ${\bm v}$ be those of ${\hat M}^0_0$ and ${\hat M}^1_0$, respectively. Formally, we have
\begin{eqnarray}
{\hat \rho}_{x_0x_1}=\frac{1}{2}(I+{\bm r}_{x_0x_1}\cdot{\hat {\bm \sigma}})\qquad(\|{\bm r}_{x_0x_1}\|\leq1),
\end{eqnarray}
\begin{eqnarray}
{\hat M}^0_t=\frac{1}{2}(I+(-1)^t{\bm u}\cdot{\hat {\bm \sigma}}),
\end{eqnarray}
and
\begin{eqnarray}
{\hat M}^1_t=\frac{1}{2}(I+(-1)^t{\bm v}\cdot{\hat {\bm \sigma}}),
\end{eqnarray}
where ${\hat {\bm \sigma}}=({\hat \sigma}_x,{\hat \sigma}_y,{\hat \sigma}_z)$. The optimality of the measurement implies that $\|{\bm u}\|=\|{\bm v}\|=1$. From the condition (\ref{eq:bloch1}) and (\ref{eq:bloch2}), we obtain
\begin{eqnarray}
{\bm u}\cdot{\bm r}_{00}={\bm u}\cdot{\bm r}_{01}=-{\bm u}\cdot{\bm r}_{10}=-{\bm u}\cdot{\bm r}_{11}=\alpha\;,\nonumber\\
{\bm v}\cdot{\bm r}_{00}=-{\bm v}\cdot{\bm r}_{01}={\bm v}\cdot{\bm r}_{10}=-{\bm v}\cdot{\bm r}_{11}=\beta\;.\;
\end{eqnarray}
Let ${\bar{\bm r}}_{x_0x_1}$ be the projection vectors of ${\bm r}_{x_0x_1}$ onto the two dimensional subspace spanned by ${\bm u}$ and ${\bm v}$. Then we have
\begin{eqnarray}
{\bar{\bm r}}_{00}+{\bar{\bm r}}_{11}={\bar{\bm r}}_{01}+{\bar{\bm r}}_{10}={\bm 0}\;.
\end{eqnarray}
and
\begin{eqnarray}
{\bm u}\cdot({\bar{\bm r}}_{00}-{\bar{\bm r}}_{01})={\bm v}\cdot({\bar{\bm r}}_{00}-{\bar{\bm r}}_{10})=0\;.
\end{eqnarray}
Due to the optimality of the decoding measurements, we also have ${\bm u}\parallel({\bar{\bm r}}_{00}+{\bar{\bm r}}_{01})$ and ${\bm v}\parallel({\bar{\bm r}}_{00}+{\bar{\bm r}}_{10})$. Thus we obtain ${\bm u}\cdot{\bm v}=0$. Hence
\begin{eqnarray}
\alpha^2+\beta^2=({\bm u}\cdot{\bar{\bm r}}_{x_0x_1})^2+({\bm v}\cdot{\bar{\bm r}}_{x_0x_1})^2\leq\|{\bm r}_{x_0x_1}\|^2\leq1\;.
\end{eqnarray}

\section{Inclusion relation of the sets of no-signalling correlations} 
Inclusion relations of the sets of bipartite and multipartite no-signalling correlations are given in (\ref{hierarchy}). 

\begin{eqnarray}
\mathcal{NS}=\mathcal{NSS}\supset\mathcal{IC}\supseteq\mathcal{CR}\supseteq\mathcal{Q}\supset\mathcal{C}\label{hierarchy}\\
\hspace{7mm}(a)\hspace{10mm}(b)\hspace{6mm}( c)\hspace{6mm}(d)\hspace{4mm}(e)\nonumber
\end{eqnarray}

$\mathcal{NS}$ is the set of all no-signalling correlations. $\mathcal{NSS}$ is the set of all no-signalling correlations that satisfies the no-supersignalling condition. By ``satisfy'' we mean that for any communication protocol using that correlation, the condition is never violated. Similarly, $\mathcal{IC}$ and $\mathcal{CR}$ are the sets of all no-signalling correlations that satisfy information causality and the chain rule, respectively. $\mathcal{Q}$ and $\mathcal{C}$ are the sets of quantum and classical correlations, respectively. $\supset$ represents the {strict} inclusion relation, and $\supseteq$ indicates that we do not know whether the sets are equivalent or {strictly included}. $(a)$ is proved in Section \ref{sec:cr}. $(b)$ is proved in \cite{ic1}. $( c)$ follows from the discussion in Section \ref{sec:ex}. $(d)$ is obvious and $(e)$ is proved in \cite{bell}. Recently it is proved from the observation of tripartite nonlocal correlations that at least one of $( c)$ and $(d)$ is a {strict inclusion} \cite{multi1, multi2}.


\end{document}